# Electro-optic Controllable Disordered Photonic Crystal for Anderson Localization of Light in Real Time

Suman Kalyan Manna[1,2*], Laurent Dupont[1], Sasanka Dalapati[3*]

[1]Department of Optics, Telecom Bretagne, 655 Avenue du Technopole, 29200 Plouzané, France
[2]Department of Cell Biology, University of California, Davis, 95616, United States
[3]Department of Chemistry, Indian Institute of Engineering Science and Technology, Shibpur, 711103, India

**Anderson localization is a ubiquitous interference phenomenon in which waves fail to propagate in a disordered medium. Unlike in a classical resonator, satisfying the favorable condition for the interference in a disordered medium is truly a statistical problem in physics. Recent progress in realizing Anderson localization is mainly limited to the iterative method for optimizing the disordered medium. Availability of an *in-situ*, active control for optimization surely paves the way of realizing the Anderson localization and its applications. In this letter, we have proposed an electro-optic controllable disordered photonic crystal and demonstrated its performance in terms of Anderson localization of light in situ in real time by application of an external electric field. We believe that Anderson localization using this medium is not only expected to address the scientific rigors but also to introduce an extra degree of freedom i.e., the tunability in its technical applications.**

The statistical distribution of phasors for maximizing the possibility of constructive interference unequivocally leads to a criterion, known as Ioffe-Regel criterion[1,2], $k\ell_s \sim 1$, for Anderson localization, where $k = 2\pi/\lambda$, and $\ell_s$ is the scattering mean free path, defined by the average distance between any two consecutive scattering events. In the dilute limit of scatterer, the optical density, $1/\ell_s$ increases linearly with the particle volume fraction $\emptyset$, i.e., $1/\ell_s \propto \emptyset$. It is observed that[3,4], above a certain value of $\emptyset$, short-range interparticle correlation starts prevailing in the system and instead, results decreased in the optical density $1/\ell_s$. Intuitively, there should be a region between the uncorrelated (dilute limit) and correlated (high concentration limit) regimes where the value of $\ell_s$ must reach to a global minimum. Searching this critical regime has been an emergence over the last few decades to demonstrate the Anderson localization of light[5-9]. Because, the Ioffe-Regel criterion is likely to be held at this regime. Parameterizing this region of interest requires the knowledge of the local structure factor $S(q)$ of the disordered medium as well as the nature of the constrains, such as, absorption coefficient, coherent enhancement factor, internal reflection, etc. Note that, unless there is an *in-situ* control, optimizing all these entities experimentally, requires a multiple iterative step. This limits the adaptivity and reproducibility of this localization technique for various photonic applications. An ability to tune the disorder or in other word, searching the effective true-statistics for the constructive interference *in-situ*, in real-time must be a privilege to define the localization regime even without solving the complicated light diffusion equations. In this letter, we introduce an *in-situ* control of disorder, for the first time for realizing Anderson localization of light, by application of an external electrical field in liquid crystal (LC) based disordered photonic crystal and validate its ability in terms of the Anderson localization of light.

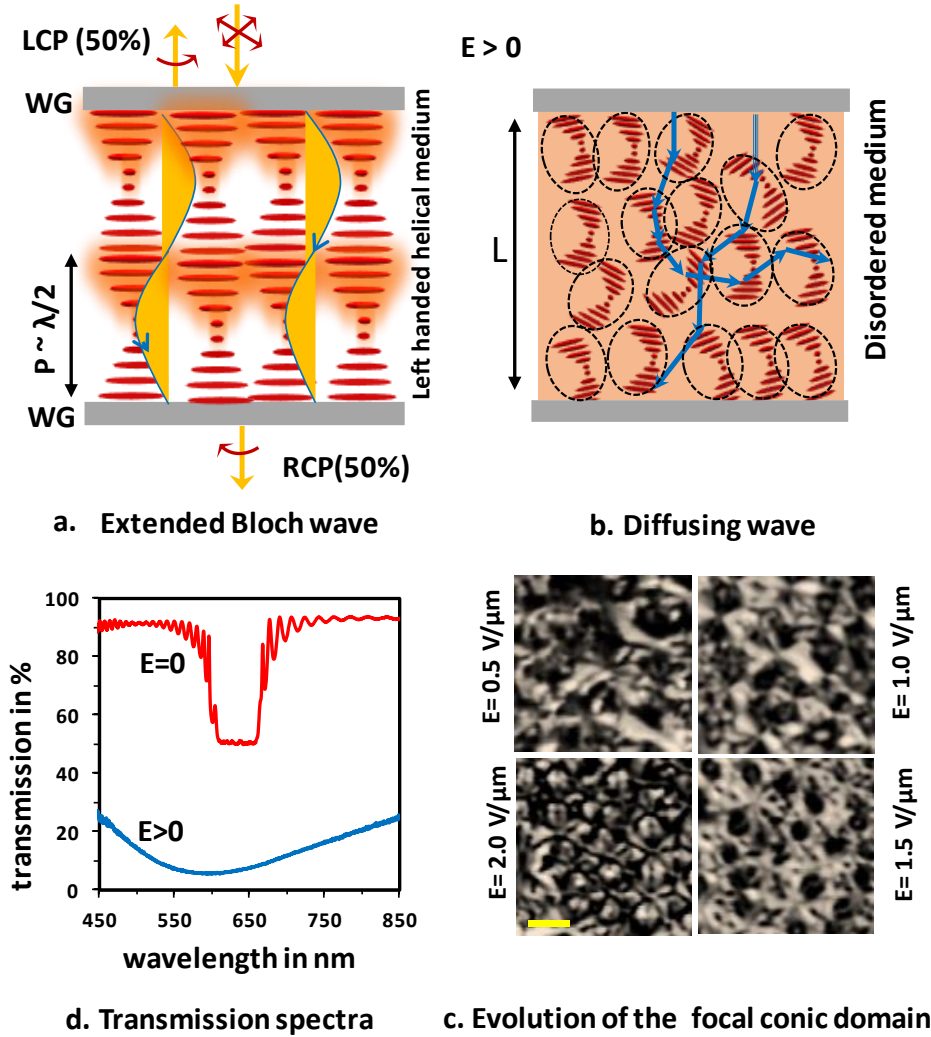

**Fig.1.Chiral-nematic liquid crystalline mesophase. a.** Schematic diagram of the self-assembled periodic dielectric layers, which support extended Bloch wave; P: pitch length of the structural helix; WG: window glass to encapsulate the liquid crystal sample; LCP: left circular polarization; RCP: right circular polarization. The chiral-nematic liquid crystal medium is shown here to be a left-handed helix hence, polarization selectivity property of this medium renders RCP to transmit through the film and LCP to reflect back upon shining an un-polarized light on it. **b.** External electric field induced distorted focal conic state is shown schematically, in which multiple light scattering can be expected from the random domains; L: thickness of the cell. **c.** A set of polarizing optical microscopic images (illumination at 488nm using a lens of focal length $f = 5mm, NA = 0.41$, acquired from a single cell of thickness L=5µm, containing 25wt% -R811, a chiral dopant mixed in a host liquid crystal -E44), shows how the focal conic domain sizes are evolved with application of an external electric field (values are shown on the respective images); scale bar $5\mu m$. **d.** Transmission spectra of the same cell reveals photonic band gap (RCP 50%) at field E= 0, where the shape of the transmission spectrum is changed with increasing the field value, E> 0, and the overall transmission is decreased because of the multiple light scattering from random focal-conic domains (as in Fig.1.b.).

LC has been a very promising material for tunable optics since its glorious historical beginning. Particularly, its electro-optic (e-o) control and self-assembled property made the material highly efficient for a large range of applications including display, laser, filter, wave-plates, lens, grating, polarizer, photonic crystals and more[10]. The advantage of its self-assembled property[11] is not only limited to reducing the fabrication cost for various photonic applications, but also, likely to enhance the monodispersity, and reproducibility of a certain structure factor $S(q)$. The self-assembled structural analysis for various mesophases of LC and their e-o properties have been well studied[12,13]. Of interest a mesophase, called chiral-nematic, prepared by mixing a solvable chiral dopant with a nematic LC is treated as self-assembled, soft-matter photonic crystals. A complete photonic band gap is appeared in the visible spectral range due to the selective Bragg's reflection from a set of self-assembled periodic dielectric layers (Fig.1.a.). Being e-o material, the self-assembled structure factor $S(q)$ can easily be deformed (Fig. 1.c.) by application of an external electric field[14-16]. The deformed state is known as focal conic state (Fig.1.b.), which scatters light strongly. Because, multiple light scattering is occurred from a set of random domains appeared at this state[16]. Here, we intend to exploit this electric field controlled focal conic state towards realization of the Anderson localization of light. For that, we optimized the scattering strength of the focal conic state by varying the chiral concentration (Fig.2.a) and, electric field (Fig.2.b). We use this optimized mixture in the following section for quantifying the localization of light.

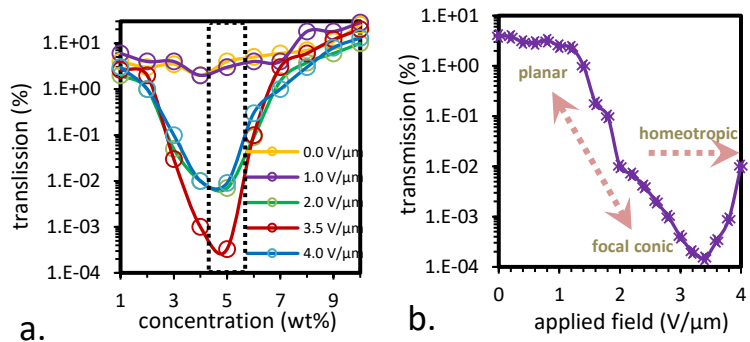

Fig.2. Optimization of chiral-nematic mixture. a. a few demonstrative curves showing the variation of transmission T (in %) in the semi-log plot by varying the chiral-concentrations (in wt%) for five different fields (cell thickness 20μm). b. semi-log plot of the variation of T with externally applied electric field, only for the optimized mixture (having chiral-concentration ~5.2wt% shown by a rectangular box in Fig. 2.a.).

Scaling theory of the localization predicts[17] that the transmission $T$ decays exponentially, i.e., $T \propto \exp(-L)$ with the thickness of the sample $L$ within the localization regime ($k\ell_s \sim 1$). Whereas, the decay shows quadratic nature i.e., $T \propto L^{-2}$ outside the regime. We use these predictions as one of the approaches, likely to confirm the Anderson localization of light within our proposed e-o controllable disordered photonic crystals. Fig.3.a. shows the quadratic nature of the transmission over inverse-thickness of the sample (optimized one, ~5.2wt% of chiral concentration) for a few demonstrative field-values. Here, the quadratic nature suggests that these corresponding filed-induced (1.5V/$\mu m$ and 2.9V/$\mu m$) disorders are not sufficient for light-localization but, light-diffusion. The characteristic exponential type decay behavior is observed for 3.5V/μm (Fig.3.b.), whereas the decay follows the exponential trend with large deviations for two other near-neighboring fields (3.2V/μm and 3.8 V/μm). These collective results apparently show the ability to tune the disorder of the soft-matter photonic crystal by application of external electric field. Secondly, the exponential decay of transmission for the typical field-value (3.5V/μm) is likely to satisfy the criterion for Anderson localization. Comprehensively, at this optimized field-value (E$_{opt}$ = 3.5V/μm), the structure factor, $S_q(E_{opt})$ of the deformable domain becomes highly favorable for maximizing the constructive interference among its statistically distributed random phasors: the realization of true-statistics for localization.

The rate of the exponential decay of transmission is designated as the localization length ($l_{loc}$). Fig.3. b. shows a comparison between the rates of the decay at three different fields. For $E_{opt}$ =3.5V/μm, the value of $l_{loc}$ gets shorter (~2.4μm). We image a bright spot appeared for these three field values (Fig.3.c). The 3D surface profile provides the spatial distributions of intensity for comparison (Fig.3.d.). It is observed that the tail of the distribution gets faster decay for $E_{opt}$ =3.5V/μm. This in turn indicates (from the central limit theorem) that the associated number-density of independent phasors (or, summands) is higher. In other word the chance of happening the true-statistics is higher at $E_{opt}$, whereas the chance is reduced gradually while moving away bi-directionally from this optimize field. Note that the reason of this bi-directional reduction is due to the e-o property of this material which is discussed in the **supplementary** (S1).

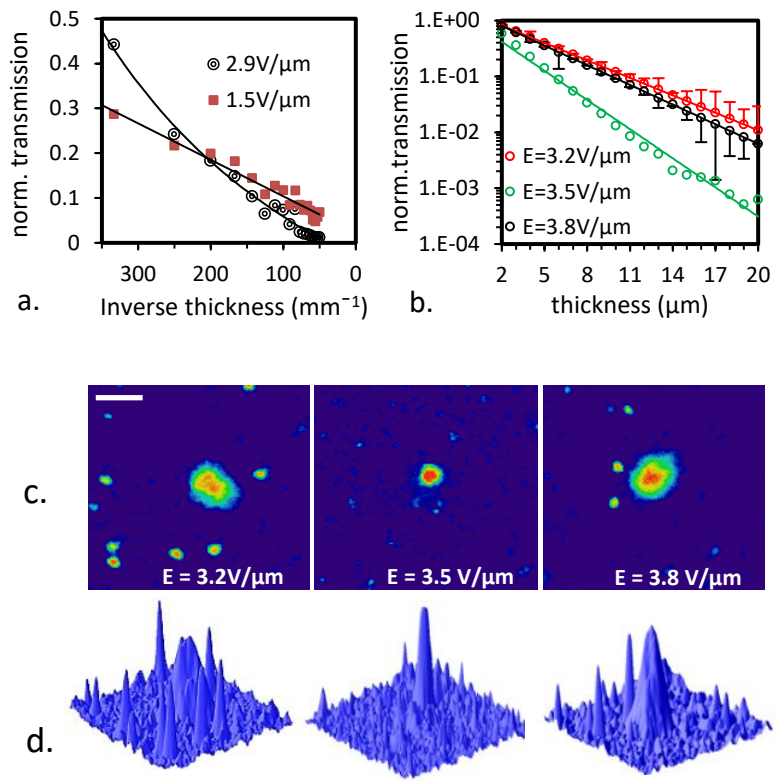

Fig.3. Quantification of the Anderson localization for the optimized mixture (5wt% chiral concentration). **a. & b.** Transmission based study shows progressive change in the decay curve with various fields from linear at 1.5V/μm to exponential at 3.5 V/μm (**b.**) through the quadratic nature around at 2.9V/μm. The rate of the decay i.e., the localization length which is obtained to be ~2.4 from the exponential fit $T \propto \exp(-L/l_{loc})$ to the curve corresponds to the optimized field $E_{opt} = 3.5$V/μm. **c.** A sequence of images (illumination $\lambda$ =488nm, objective lens $f = 5mm, NA = 0.41$) are taken to compare the evolution of the localized spot and, **d.** their spatial-distributions in intensity.

Another notable approach for identifying the prerequisite of Anderson localization is performed here based on the far-field speckle statistical. The detailed development of the statistical analysis can be found elsewhere[18]. Briefly, the turbidity in the optical phase near to the localization regime is higher which reflects to the statistical property of the transmitted light in such a way that, the far-field intensity fluctuation becomes higher. Here, we measure the far-field transmitted intensity (Fig.3.a.) along a line (line profile) parallel to the LC cell surface for various fields (see the **supplementary** for the setup). Two statistical parameters, the variance $\sigma_I$, and the mean $\langle I \rangle$ for the intensity distributions are estimated for comparison.

The value of $\langle I \rangle$ is observed to be less for $E_{opt}$ which is expected. Because ideally, the light is localized within the sample at $E_{opt}$. Further, the value of $\sigma_I$ is found to be higher at $E_{opt}$, which is also expected as per the statistical signature of the localization.

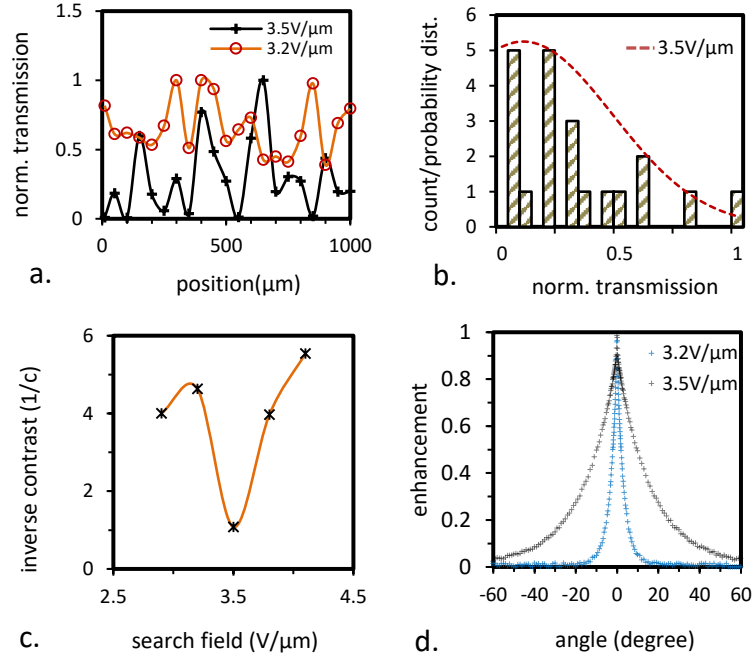

**Fig.4. Evaluation of the far-field speckle statistical. a.** The statistics of the far-field transmitted intensity along a line parallel to the LC cell surface showing higher contrast $c = \sigma_I/\langle I \rangle = 0.25$, for $E_{opt}$ and, 0.06 for a near-neighboring search field 3.2V/µm. **b.** The histogram and normal distribution of the far-field transmission are plotted only for $E_{opt}$. **c.** The inverse contrast $1/c = \langle I \rangle/\sigma_I$ of the transmission profile for each field is plotted. **d.** Finally, measurement of the turbidity is estimated from the coherent backscattering to derive $\ell_s$, which is ~150nm for $E_{opt}$.

Finally, evaluation of optical turbidity is done through measuring the coherent backscattering for $E_{opt}$ as well as, for a near neighboring electric field, 3.2V/$\mu m$(Fig.4.d.). The detailed geometry of the system is illustrated in **supplementary**. The values are found to be $k\ell_s \sim 1.9$ for $E_{opt}$ and, $\sim 7.2$ for the near neighboring field by measuring the inverse width of the backscattering cone fitting the low angle slope in the region of the strong scattering[19]. Note that, the minimum we obtained ($k\ell_s \sim 1.9$) from our experiment is surely off-centered from the global minimum ($k\ell_s \sim 1$), hypothesized for the Anderson localization of light. However, this difference lies well inside the range of value reported in literature, so far[5-9]. A few general reasons mainly absorption loss[20-24], internal reflections are summoned most often to play the pivotal role towards this difference. In our case, these possible reasons may be applicable as well. Moreover, from statistical point of view, any disordered system always possesses some non-identical distributions, which limit the system form generating the true-statistics. In our case, the subjective origin of these non-identical distributions may be listed as inhomogeneity in mixture, phase segregation, non-uniformity in sample thickness and the corresponding dielectric coupling etc. Presence of these most probable experimental factors can oppose to happen the true-statistical distribution of the phasors, even in our fine e-o controllable disordered photonic crystal.

In conclusion, we introduce an electric field controllable disordered photonic crystal for realizing Anderson localization of light. The extent of the localization is analyzed from multiple approaches; firstly, scaling theory-based approach, which predicts exponential decay of transmission with sample-thickness within the localization regime. Secondly, we use the central limit theorem in statistics for analyzing the image of the localized spots, followed by the far-field speckle statistics for the transmitted intensity and, finally, the measurement of coherent backscattering. We found that the Ioffe-Regel criterion, $k\ell_s \sim 1$, for localization is closely held in our proposed e-o controllable disordered photonic crystal. We believe that Anderson localization using this material is not only expected to address the scientific rigors but also to introduce an extra degree of freedom i.e., the tunability in its technical applications.

**Method:**

**Sample preparation:** For chiral-nematic mixture, we used chiral dopant R811 (HTP $\sim 11\ \mu m^{-1}$; Merck) mixed with a nematic liquid crystal E44 ($n_{avg} = 1.54$, $T_{NI} \approx 100°C$, Merck).

**Cell fabrication:** All the liquid crystal cells used here are custom designed. There is no alignment layer used on the glass substrates in order to exploit the fullest randomness of the director orientation of liquid crystals. Maintaining and evaluating thickness of the cell is very important for characteristic transmission vs. thickness plot. In order to maintain the uniformity of the thickness, spacer particles (of required thickness) are dispersed (0.1wt%) in isopropanol and the mixture solution is spin coated at 3000rpm on the glass substrates in order to have an uniform distribution. After spin coating, the substrates kept in an oven at 150 °C for 30 mints to evaporate the isopropanol. These substrates are then used to make a cell within which the chiral-nematic LC mixture is infiltrated. Note that, all the mixtures are infiltrated at the isotropic temperature of the liquid crystal (~113°C) to maintain the homogeneity of the mixture. The effective thickness $L$ of a mixture sample is determined from the Fabry-Perot interference fringes of the cell after filling the mixture.

**Driving field:** It is instructive to apply continuous RF wave to a liquid crystal cell in order to avoid the influence of ionic current[12,13] inside the cell which can damage the cell and liquid crystals. Because of that we applied continuous FR wave with frequency 1KHz. The value of this frequency kept constant during the entire experiment. For transmission vs. applied field measurement (Fig.2b), the reading is taken during increment of the field gradually. This trend is in fact, followed during the entire study in order to avoid any hysteresis (memory) effect[14] related to the reverse voltage scanning.

**Acknowledgements**

This research was supported by TVSF collaborative research grant for Integrated Photonics research and Eu NoE Nanophotonics for energy efficiency. We would like to thank to Prof. H. Folliot, and K. Sathaye for support and fruitful discussion.